\documentclass[preprint,tighten,aps,epsfig]{revtex4}

\newcommand{\bea}{\begin{eqnarray}}
\newcommand{\eea}{\end{eqnarray}}

\newcommand{\lp}{\left}
\newcommand{\rp}{\right}
\begin{document}
\preprint{JLAB-THY-01-35}
\title{Masses of the $70^-$ Baryons in Large $N_c$ QCD}
\author{
C. L. Schat $^{a}\ ^\dagger$ \thanks{e-mail: schat@jlab.org}, \
J. L. Goity $^{a,b}$ \thanks{e-mail: goity@jlab.org}, \
N. N. Scoccola $^{c,d}$ \footnote[2]{Fellow of CONICET, Argentina.} \thanks{e-mail: scoccola@tandar.cnea.gov.ar}}

\affiliation{
$^a$ Thomas Jefferson National Accelerator Facility, Newport News, VA 23606, USA. \\
$^b$ Department of Physics, Hampton University, Hampton, VA 23668, USA. \\
$^c$ Physics Department, Comisi\'on Nacional de Energ\'{\i}a At\'omica, Av.~Libertador 8250, (1429) Buenos Aires,
     Argentina.\\
$^d$ Universidad Favaloro, Sol{\'\i}s 453, (1078) Buenos Aires, Argentina.}

\date{\today}
\begin{abstract}
The masses of  the negative parity 70-plet  baryons are analyzed in  large $N_c$ QCD  to order  $1/N_c$
and to first order in $SU(3)$ symmetry breaking. The existing experimental data are  well reproduced and twenty new
observables are predicted. The leading order $SU(6)$ spin-flavor symmetry breaking is  small and, as it occurs in
the quark model, the subleading in $1/N_c$ hyperfine interaction is the dominant  source of the breaking.
It is found  that the $\Lambda(1405)$ and  $\Lambda(1520)$  are  well described as three-quark states and
spin-orbit partners. New relations between  splittings in different $SU(3)$ multiplets are found.
\end{abstract}

\maketitle

The study of excited mesons and baryons has been largely the domain of the quark model \cite{CandR}. Despite the
success of this model in reproducing  the general features of the spectrum and decays, it is clear that in its
different versions it is not a complete representation of QCD. One consequence of this incompleteness is that in
those cases where the quark model does not agree with phenomenology, such as the problem of the mass splittings between
spin-orbit partners in the negative parity baryons (spin-orbit puzzle), it is not clear whether the problem is due
to the quark model itself or to specific dynamical properties of the states involved. In the last few years it was
realized that the $1/N_c$ expansion can provide a  link between the phenomenology of excited baryons and QCD that
avoids the assumptions required in the  quark model. This link has the form of an effective theory that implements
an expansion in $1/N_c$. As shown in this Letter, the main features of the quark model emerge unscathed from the
large $N_c$ analysis of the masses in the 70-plet of negative parity baryons, and in addition some of the missing
pieces of the model are recovered. For example, the long standing spin-orbit puzzle seems to be easily resolved by
the presence of one operator of ${\cal{O}}(N_c^0)$ not included in the quark model.

In the $N_c \rightarrow \infty$ limit of QCD
the ground state baryons display a contracted dynamical spin-flavor
symmetry $SU(2F)$ ($F$ is the number of flavors, equal to three in this Letter), which is a
 consequence of unitarity in  pion-nucleon scattering in that limit
\cite{GeSa84,Dashen1}. In general $SU(2F)$  is broken at ${\cal O}(1/N_c)$ and
for some observables even at ${\cal O}(1/N_c^2)$ \cite{Dashen1}. This implies  that
perturbation theory around  the  $SU(2F)$ symmetric limit in the form of a $1/N_c$ expansion
is a powerful tool of analysis, as shown in numerous works \cite{Dashen1,CGO94,LM94,JL95,Man98}.
In the context of the $1/N_c$ expansion the sector of excited baryons
\cite{CGKM94,Goi97,PY98,CC00} is less understood.
The principal reason is that even in the $N_c \rightarrow \infty$ limit there is no exact dynamical symmetry \cite{Goi97}. However, an important simplification results from the  observation that excited states can be classified
into multiplets of spin-flavor $SU(2 F)$.
 For example, most of the known baryons of negative parity
seem to fit very well in the $(3,70)$ irreducible representation (irrep) of $O(3)\otimes SU(6)$. In particular, this implies that the analysis of  excited baryon masses can be carried out within a spin-flavor multiplet  along the lines recently developed for two flavors \cite{Goi97,CCGL}.

The  states in the $(3, 70)$ decompose in terms of  $ SU(2)\otimes SU(3)$ into two octets with total angular
momentum $J=1/2$ ($^{2 S+1}d=$ $^28$ and $^48$, where $S$ is the total spin and $d$ the degeneracy of the $SU(3)$
irrep), two octets with $J=3/2$ ($^28$ and $^48$), one octet with $J=5/2$ ($^48$), one decuplet with $J=1/2$ and
one  with $J=3/2$ (both $^210$), and two singlet $\Lambda$s with $J=1/2$ and $3/2$ (both $^21$).

Since in the large $N_c$ limit baryons consist only of valence quarks, it is natural to have an intuitive
non-relativistic quark model picture of the spin-flavor composition of the states. This only means that the
identification of spin-flavor states in the large $N_c$ analysis and the quark model is the same. Thus, the wave
functions are  constructed by coupling an orbitally excited quark with $\ell=1$ to $N_c-1$ s-wave quarks that are
in a spin-flavor  symmetric core. The states have the general form
\bea
|\Psi> &=& |J,J_z ; S ; (\lambda,\mu), Y,I,I_z> \nonumber \\
&=& \sum_{\alpha, \alpha', \alpha''} {\cal CG}_{\alpha, \alpha', \alpha''}  |l>_{\alpha} |q>_{\alpha'} |c>_{\alpha''}
\ ,
\eea
where $\alpha$  stands for the different projection quantum numbers and ${\cal CG}$ for
Clebsch-Gordan coefficients. In addition, the $(\lambda,\mu)$ labels indicate the $SU(3)$ irrep, $Y$ is the
hypercharge, $I$ the isospin and $J_z$, $I_z$ the obvious projections.
The (3,70) states  are taken  to have strangeness of order $N_c^0$.
The excited quark and  core states are given  in terms of their $SU(2)\otimes SU(3)$ quantum numbers with
obvious notation:
\bea
| q > =  \lp| \begin{array}{cc}
\frac{1}{2}  & (1 , 0)  \\
s_z & (y ,  \frac{1}{2}  , i_z )
\end{array}
\rp\rangle  &,&
|c> =
\lp|
\begin{array}{cc}
S^{c}   & (\lambda^{c} , \mu^{c})  \\
S^{c}_z & (Y^c, I^c, I_z^c)
\end{array}
\rp\rangle ,
\eea
where  $S^c$ is the spin of the core. From the decomposition of the $SU(6)$ symmetric
representation into representations of $SU(2) \otimes SU(3)$  the relations $\lambda^c + 2 \mu^c = N_c -1$ and
$\lambda^c = 2 S^c$ follow. They are the generalization of the $I=J$ rule well known for two flavors.
The total wave function is in the mixed symmetric irrep of $SU(6)$. In
the  $^2 8$ representation  a linear combination of two core states appears, namely $|S^c, (\lambda^c,\mu^c)> =
|0,(0,\frac{N_c-1}{2})>$ and $|1,(2,\frac{N_c-3}{2})>$, while in the  $^4 8$ and $^2 10$ representations the core
state is $|1,(2,\frac{N_c-3}{2})>$,  and finally,  in the $^2 1$ representation  the core state is $
|0,(0,\frac{N_c-1}{2})>$.

A basis of mass operators can be built using the generators of $O(3) \otimes SU(2 F )$ \cite{Goi97}.
A generic $n$-body mass operator has the general structure
\bea
O^{(n)} &=& \frac{1}{N_c^{n-1}} \ O_{\ell} \ O_q \ O_c \ ,
\eea
where the factors $O_{\ell}$, $ O_q$, and  $O_c$ can be expressed in terms of products of generators of
orbital angular momentum ($\ell_i)$,  spin-flavor of the excited quark ($s_i, t_a$ and $g_{ia} \equiv s_i t_a$) and
spin-flavor of the core ($S^c_i, T^c_a$ and $G^c_{ia} \equiv \sum_{m=1}^{N_c-1} s^{(m)}_i t^{(m)}_a$), respectively.
The explicit $1/N_c$ factors originate in the $n-1$ gluon exchanges required to
give rise to a $n$-body operator.
The matrix elements of operators may also carry a nontrivial $N_c$ dependence due to coherence
effects \cite{GeSa84,Dashen1}:
for the states considered, $G^c_{ia}$ ($a=1,2,3$) and $T^c_8$  have matrix elements of  ${\cal{O}}(N_c)$,
while  the rest of the generators have matrix elements of higher order.

For $N_c=3$ and in the $SU(3)$ limit there are eleven independent quantities, namely nine masses and two mixing
angles $\theta_1$ and $\theta_3$, which correspond to the mixing of the $^2 8$ and $^4 8$
irreps in the $J=1/2$ and $J=3/2$ octets, respectively.
There is, therefore, a basis of eleven $SU(3)$-singlet
mass operators.
As shown in Table I, the basis of singlet operators $O_i$ consists of  one operator of ${\cal O}(N_c)$, namely the identity
operator, three  operators of ${\cal O}(N_c^0)$, that include the spin-orbit operator, and seven of ${\cal O}(1/N_c)$,
one of which is the very important hyperfine operator. These operators are a simple generalization of
those known for two flavors
\cite{footdos}.

When $SU(3)$ breaking is included with  isospin conservation, the number of independent
observables raises up to fifty, of which thirty are masses and twenty are
mixing angles.  However, if $SU(3)$ symmetry breaking is restricted to linear order in quark masses
 only isosinglet octet
operators can appear, and the number  of independent observables is reduced to thirty five (twenty one masses and
fourteen mixing angles) implying  twenty four linearly independent octet mass operators.
As a consequence of this reduction several mass relations exist, among them there is a Gell-Mann Okubo relation for each
octet and an equal spacing rule for each decuplet.
The octet contributions are proportional to
$\ \epsilon~\propto~(m_s~-~m_{u,d}~)/\nu_{H}$ where $\nu_{H}$ is a typical
hadronic mass scale, for instance $m_\rho$;  for $N_c=3$ the quantity $\epsilon$ counts as of the same order as
$1/N_c$. Explicit construction shows that up to order ${\cal O}(\epsilon N_c^0)$ only a small subset
of independent octet operators $B_i$ appears. Since such octet operators are isospin singlets, it is possible to modify
them by adding singlet operators so that the resulting operators vanish in the subspace of
non-strange baryons. This procedure of improving the flavor breaking operators may change the $1/N_c$ counting: for
instance, after improving $T_8$ with the identity operator $O_1$ the resulting operator is of order $N_c^0$. Indeed, the
improved operators give the splitting due to $SU(3)$ breaking with respect to the non-strange baryons in each
multiplet, and they must be of zeroth order  or higher in $1/N_c$ for states with strangeness of order $N_c^0$.
The  four improved flavor breaking operators $\bar B_1$ through $\bar B_4$ that remain
at ${\cal O}(\epsilon N_c^0)$ when $N_c=3$ are shown in Table I.

As a result of the  above analysis  the 70-plet  mass operator up to  ${\cal O}(1/N_c)$
and ${\cal O}(\epsilon N_c^0)$  has the most general form:
\bea
M_{70} & =& \sum_{i=1}^{11} c_i O_i + \sum_{i=1}^{4} d_i \bar B_i~~~~,
\eea
where $c_i$ and $d_i$ are numerical coefficients which can be determined by fitting
the available empirical masses and mixing angles. For this purpose it is necessary to
have  the expressions of the matrix elements
of the $O_i$ and $\bar B_i$ operators between a basis of states belonging to the 70-plet.
Their analytic expressions are  obtained using standard techniques and will be  given elsewhere.

Because at 
${\cal O}(\epsilon N_c^0)$ there are only four flavor breaking operators,
 it is possible to find  new mass splitting relations which are independent of the coefficients $d_i$. These 
relations involve states in different $SU(3)$ multiplets.
Of particular interest are the following five relations that result when the
operator $\bar B_3$ is neglected (from the fit below it is apparent that $\bar B_3$ gives very small contributions):
\bea
9 (s_{\Sigma_{1/2}} +  s_{\Sigma'_{1/2}}) + 21 s_{\Lambda_{5/2}} &=&
17 (s_{\Lambda_{1/2}} +  s_{\Lambda'_{1/2}}) + 5 s_{\Sigma_{5/2}} \ , \nonumber \\
2 (s_{\Lambda_{3/2}} +  s_{\Lambda'_{3/2}}) &=&
3  s_{\Lambda_{5/2}} + s_{\Sigma_{5/2}} \ , \nonumber \\
18 (s_{\Sigma_{3/2}} + s_{\Sigma'_{3/2}}) +
33 s_{\Lambda_{5/2}} &=& 28 (s_{\Lambda_{1/2}} + s_{\Lambda'_{1/2}})
+ 13 s_{\Sigma_{5/2}} \nonumber \ , \\
9 s_{\Sigma''_{1/2}} &=&  s_{\Lambda_{1/2}} +   s_{\Lambda'_{1/2}}
+ 3 s_{\Lambda_{5/2}} + 4 s_{\Sigma_{5/2}} \nonumber  \ , \\
18 s_{\Sigma''_{3/2}} + 3 s_{\Lambda_{5/2}} &=& 8(s_{\Lambda_{1/2}}
+s_{\Lambda'_{1/2}})  + 5 s_{\Sigma_{5/2}} \ .
\eea
Here $s_{{\cal B}_i}$ is the mass splitting between the baryon
${\cal B}_i$ and the non-strange baryons in the $SU(3)$ multiplet
to which it belongs. These relations are independent of mixings because they result from relations among traces of
the octet operators. If $\bar B_3$ is not neglected there are instead four relations.

{\it Discussion of the fit and conclusions  -- } The experimental masses shown in Table II (three or
more stars status in the the Particle Data listing \cite{PDG}) together with the two leading order
mixing angles $\theta_1= 0.61$,  $\theta_3= 3.04$ \cite{HLC75,ik78} are the 19 empirical quantities to be
fitted. One three-star state is not included, namely the $\Sigma(1940)$ which does not consistently
fit into the 70-plet, and the two-star states  $\Sigma(1580)$ and $\Sigma(1620)$ are not included as
inputs. The errors in mass inputs are taken to be equal to  the experimental errors if these  are larger than
the  magnitude of the theoretical errors  estimated at $\pm 15 \;{\rm MeV}$, otherwise they are taken to
be equal to the latter. The fifteen coefficients $c_i$, $d_i$ are obtained from the fit, and the
resulting  $\chi^2$  per degree of freedom of the fit  turns out to be $\chi^2 /4 = 1.29$. The results
for the coefficients are displayed in Table I, while the best fit masses and state compositions are
displayed in Table II. Note that the natural size of coefficients associated with singlet operators is
set by the coefficient of $O_1$, and is about  $500 \ {\rm MeV}$ , while the natural size for the
coefficients associated with octet operators is roughly $ \epsilon $ times $ 500  \ {\rm MeV}  $.

There are a number of important  points that emerge from this analysis.

Although spin flavor symmetry  is broken  at  ${\cal O}(N_c^0)$, it is evident that the ${\cal O}(N_c^0)$ operators
are dynamically suppressed as their coefficients are substantially smaller than  the natural size. It turns out that
the chief contribution to spin-flavor breaking stems from  the ${\cal O}(1/N_c)$ hyperfine operator  $O_6$, as in
 the ground state baryons. Since $O_6$ is purely a core operator, it turns out that the gross
spin-flavor structure of levels is determined by the two possible  core states. This observation is in agreement
with the findings of  quark models \cite{ik78,GR96}. In particular, the two singlet $\Lambda$s are not affected by $O_6$, while  the other
states are moved upwards, explaining in a transparent way the lightness of these two states. Indeed, by keeping
only $O_1$ and $O_6$ the $^2 8$  masses are $1510 \ {\rm MeV}$, the $^4 8$ and $^2 10$ masses are $1670 \ {\rm
MeV}$, and the $^2 1$ masses are left at the bottom with $1350  \ {\rm MeV}$. This clearly  shows  the dominant
pattern of spin-flavor breaking observed in the 70-plet.

The  long standing  problem in the quark model  of  reconciling the large $\Lambda(1520)-\Lambda(1405)$ splitting
with the  splittings between  the other spin-orbit partners in the 70-plet is resolved  in the large $N_c$ analysis.
The singlet $\Lambda$s  receive contributions to their masses  from $O_1$ and $\ell . s$ while the rest of the operators give vanishing contributions because  their core  has $S^c=0$. The splitting between the singlets is, therefore,  a clear display of the spin-orbit coupling. The problem with the splittings between spin-orbit partners in the non-singlet sector, illustrated  by the fact that the  $\ell . s $ operator gives  a contribution to the $\Delta_{1/2}-\Delta_{3/2} $ splitting  that is of  opposite sign
 of what is observed,  is now solved by the presence of the operators $O_4$, $O_5$, $O_9$ and $O_{11}$, with the  contribution from  $O_4$ being the dominant one in accordance with the $1/N_c$ counting. One important consequence of this result is that the interpretation of the singlet $\Lambda$s as three-quark states is consistent with the masses of the rest of the 70-plet. This further supports 
a similar claim drawn from scaling down to the strange sector the mass splitting between the $\Lambda_c(2593)$ and the  $\Lambda_c(2625)$ \cite{Isg95}.

There is a hierarchy of mixing effects. As already mentioned, at  ${\cal O}(N_c^0)$ there are two mixing angles, namely $\theta_1$ and
$\theta_3$ that mix the octets with same $J$. These mixing angles are inputs and are obtained from an analysis of
the $N^*$ decays \cite{HLC75,ik78}. All ${\cal O}(N_c^0)$ operators in principle contribute to these mixings, but  the  $\ell . s$ and  $O_4$ contributions tend to cancel each other  leaving the  $O_3$ as
the dominant one. Indeed, the coefficient of $O_3$ is largely determined by   mixing  as this operator gives only
modest contributions to the masses \cite{footuno}. The rest of the mixings are of higher order because they are due to   $SU(3)$
breaking,    and   expected to be small. Table II shows this in the composition of states. A good example is that
the $\Lambda(1405)$ and  $ \Lambda(1520)$ remain largely singlet states. In some cases, however,  due to close
degeneracy the $SU(3)$ breaking can induce a larger than expected mixing angle which cannot be predicted with precision
from an analysis of the masses alone. This occurs in  the $J=3/2$  $\Sigma$s and  $\Xi$s, where for that reason the corresponding amplitudes in Table II are shown between parentheses.

The hyperfine operator between the excited $\ell=1$ quark and
the core,  $O_7$, is suppressed with respect to $O_6$,
indicating that the hyperfine interaction is of short range  in
agreement with the quark model. The large errors that make the
coefficients compatible with zero show that the operators $O_8$
and $\bar B_{3}$ are largely  irrelevant. On the other hand,  the
collective effects  of the three-body operators $O_9$, $O_{10}$
and $O_{11}$ amount  to  mass shifts of modest magnitude  (50
MeV or less).

The first relation in equation  (5) predicts the $\Sigma_{1/2}$ to be $103 \ {\rm MeV}$ above the $N_{1/2}$, consistent with
the $\Sigma(1620)$, a two star state that is not included as input to  the fit. Each of the remaining relations makes
a similar prediction for other states but requires further experimental input to be tested.

The analysis of this Letter shows that the $1/N_c$ expansion provides a systematic approach to the spectroscopy of
the negative parity baryons.  It successfully describes the existent data and to the order considered it also leads to
numerous predictions yet to be tested. In addition to the well known Gell-Mann-Okubo  and equal spacing  relations,
new splitting relations between different multiplets that follow from the spin flavor symmetry  have been found.
The $\Lambda(1405)$ is well described as a
three quark state and the spin orbit partner of the $\Lambda(1520)$. Finally,
effective interactions that correspond to flavor quantum number exchanges,
such as the ones mediated by the operators $O_3$ and $O_4$, are apparently needed.
Although the corresponding coefficients seem to be dynamically suppressed their relevance shows up in the well
established finer effects, namely mixings and splittings between non-singlet spin-orbit partners.
These interactions are not accounted for in the standard quark model based on one gluon exchange.

This work was partially
supported by the National Science Foundation through grant
\#~PHY-9733343 (JLG),
by the Department of Energy through contract DE-AC05-84ER40150 (JLG and CLS),
by CONICET (CLS), by Fundaci\'on Balseiro (NNS),
and by   sabbatical leave support  from the
Southeastern Universities Research Association (JLG). Two  of us (JLG and CLS)
 thank  the Institut f\"ur Theoretische Physik of the University
 of Bern for the kind hospitality while part of this work was completed.

\section*{\Large Table captions}

\vspace{.5cm}

\noindent {\bf TABLE I:} Operator list and best fit coefficients.

\vspace{.5cm}

\noindent {\bf TABLE II:} Masses and spin-flavor content as predicted by the large $N_c$ analysis. Also given are
the empirical masses and those obtained in the quark model (QM) calculation of Ref. \cite{ik78}.

\pagebreak

\begin{table}[h]
\centerline{TABLE I} \vspace*{.3cm}

\begin{tabular}{llrrr}
\hline
\hline
Operator & \multicolumn{4}{c}{Fitted coef. [MeV]}\\
\hline
\hline
$O_1 = N_c \ 1 $ & $c_1 =$  & 449 & $\pm$ & 2 $\ $  \\
\hline
$O_2 = l_h \ s_h$ & $c_2 =$ & 52 & $\pm$ & 15   $\ $ \\
$O_3 = \frac{3}{N_c} \ l^{(2)}_{hk} \ g_{ha} \ G^c_{ka} $ & $c_3 =$  & 116 & $\pm$ & 44  $\ $ \\
$O_4 = \frac{4}{N_c+1} \ l_h \ t_a \ G^c_{ha}$ & $c_4 =$  & 110 & $\pm$ &  16 $\ $\\
\hline
$O_5 = \frac{1}{N_c} \ l_h \ S^c_h$ & $c_5 =$  & 74 & $\pm$ & 30 $\ $\\
$O_6 = \frac{1}{N_c} \ S^c_h \ S^c_h$ & $c_6 =$  & 480 &  $\pm$ & 15 $\ $\\
$O_7 = \frac{1}{N_c} \ s_h \ S^c_h$ & $c_7 =$ & -159 &  $\pm$ & 50 $\ $ \\
$O_8 = \frac{1}{N_c} \ l^{(2)}_{hk} s_h \ S^c_k$ & $c_8  =$  & 6  & $\pm$ &   110   $\ $\\
$O_9 = \frac{1}{N_c^2} \ l_h \ g_{ka} \{ S^c_k ,  G^c_{ha} \} $ & $c_9 =$ &  213 &  $\pm$ &  153  $\ $\\
$O_{10} = \frac{1}{N_c^2} t_a \{ S^c_h ,  G^c_{ha} \}$ & $c_{10} =$  & -168 &  $\pm$ &  56  $\ $\\
$O_{11} = \frac{1}{N_c^2} \ l_h \ g_{ha} \{ S^c_k ,  G^c_{ka} \}$ & $c_{11} =$ & -133 &  $\pm$ &  130  $\ $\\
\hline
\hline
$\bar B_1 = t_8 - \frac{1}{2 \sqrt{3} N_c} O_1$ & $d_1 =$  & -81 & $\pm$ & 36 $\ $\\
$\bar B_2 = T_8^c - \frac{N_c-1}{2 \sqrt{3} N_c } O_1 $  & $ d_2 = $  & -194 & $\pm$ & 17  $\ $\\
$\bar B_3 = \frac{1}{N_c} \  d_{8ab}  \ g_{ha} \ G^c_{hb}  + \frac{N_c^2 -9}{16 \sqrt{3} N_c^2 (N_c-1)} O_1 +$ &  & & $\ $\\
\hspace*{1cm} $+ \frac{1}{4 \sqrt{3} (N_c-1)} O_6 + \frac{1}{12 \sqrt{3}} O_7 $  & $ d_3 = $  & -150 & $\pm$ & 301  $\ $\\
$\bar B_4 = l_h \ g_{h8} - \frac{1}{2 \sqrt{3}} O_2 $ & $ d_4 = $  & -82 & $\pm$ & 57  $\ $\\
\hline \hline
\end{tabular}
\vspace{1.cm}
\end{table}

\pagebreak
\renewcommand{\arraystretch}{.9}

\begin{table}
\centerline{TABLE II} \vspace*{.3cm}
\begin{tabular}{ccccccccccc}\hline \hline
       & & \multicolumn{3}{c}{Masses [MeV]} & &  \multicolumn{4}{c}{Spin-flavor content} \\
State &\hspace*{.5cm} & Expt. & Large $ N_c$  &  QM &\hspace*{.5cm} & $^21$ & $^28$ & $^48$ & $^210$ \\
\hline
  $N_{1/2}$      & &$ 1538 \pm 18  $ & 1541  & 1490  &  &               &  0.82        &  0.57  &         \\
 $\Lambda_{1/2}$ & &$ 1670 \pm 10  $ & 1667  & 1650  &  & -0.21  &  0.90  &  0.37  &         \\
 $\Sigma_{1/2}$  & &$  (1620)      $ & 1637  & 1650  &  &               &  0.52  &  0.81  &  0.27   \\
 $\Xi_{1/2}$     & &$              $ & 1779  & 1780  &  &       &  0.85  &  0.44 &  0.29  \\
\hline
 $N_{3/2}  $     & &$ 1523 \pm 8   $ & 1532  & 1535  &  &       &-0.99   & 0.10   &         \\
 $\Lambda_{3/2}$ & &$ 1690 \pm 5   $ & 1676  & 1690  &  & 0.18  & -0.98  & 0.09   &         \\
 $\Sigma_{3/2}$  & &$ 1675 \pm 10  $ & 1667  & 1675  &  &       & -0.98  & -0.01   &-0.19 \\
 $\Xi_{3/2}$     & &$ 1823 \pm 5   $ & 1815  & 1800  &  &       & -0.98  &  0.03  & -0.19   \\
\hline
 $N'_{1/2}  $    & &$ 1660 \pm 20  $ & 1660  & 1655  &  &       & -0.57  &  0.82  &         \\
 $\Lambda'_{1/2}$& &$ 1785 \pm 65  $ & 1806  & 1800  &  & 0.10  & -0.38  & 0.92  &         \\
 $\Sigma'_{1/2}$ & &$ 1765 \pm 35  $ & 1755  & 1750  &  &       & -0.83  &  0.54  &  0.17   \\

 $\Xi'_{1/2}$    & &$              $ & 1927  & 1900  &  &       & -0.46  &  0.87  &  0.18   \\
\hline
 $N'_{3/2}  $    & &$ 1700 \pm 50  $ & 1699  & 1745  &  &       & -0.10  & -0.99  &         \\
 $\Lambda'_{3/2}$& &$              $ & 1864  & 1880  &  &0.01   & -0.09  & -0.99  &         \\
 $\Sigma'_{3/2}$ & &$              $ & 1769  & 1815  &  &       & 0.01   & (-0.57)  & (-0.82)   \\
 $\Xi'_{3/2}$    & &$              $ & 1980  & 1985  &  &       & -0.02  & (-0.57)  & (-0.82)   \\
\hline
 $N_{5/2}  $     & &$ 1678 \pm 8   $ & 1671  & 1670  &  &       &        &  1.00  &         \\
 $\Lambda_{5/2}$ & &$ 1820 \pm 10  $ & 1836  & 1815  &  &       &        &  1.00  &         \\
 $\Sigma_{5/2}$  & &$ 1775 \pm 5   $ & 1784  & 1760  &  &       &        &  1.00  &         \\
 $\Xi_{5/2}$     & &$              $ & 1974  & 1930  &  &       &        &  1.00  &         \\
\hline
 $\Delta_{1/2}$  & &$ 1645 \pm 30  $ & 1645  & 1685  &  &       &        &        &  1.00   \\
 $\Sigma''_{1/2}$  & &$              $ & 1784  & 1810  &  &       &  -0.14  &  -0.31  &  0.94   \\
 $\Xi''_{1/2}$     & &$              $ & 1922  & 1930  &  &       &  -0.14  &  -0.31  &  0.94   \\
 $\Omega_{1/2}$  & &$              $ & 2061  & 2020  &  &       &        &        &  1.00   \\
\hline
 $\Delta_{3/2}$  & &$ 1720 \pm 50  $ & 1720  & 1685  &  &       &        &        &  1.00   \\
 $\Sigma''_{3/2}$  & &$              $ & 1847  & 1805  &  &       &  -0.19  &  (-0.80)  &  (0.57)   \\
 $\Xi''_{3/2}$     & &$              $ & 1973  & 1920  &  &       &  -0.19  &  (-0.80)  &  (0.57)   \\
 $\Omega_{3/2}$    & &$              $ & 2100  & 2020  &  &       &             &             &  1.00   \\
\hline
 $\Lambda''_{1/2}$ & &$ 1407 \pm 4   $ & 1407  & 1490  &  & 0.97  & 0.23  & 0.04  &         \\
\hline
 $\Lambda''_{3/2}$ & &$ 1520 \pm 1   $ & 1520  & 1490  &  & 0.98  & 0.18  & -0.01  &         \\
\hline \hline
\end{tabular}
\end{table}


\begin{thebibliography}{99}
\bibitem{footdos}
Note that $O_3$ used here is three times the one used in \cite{CCGL}.
The reason for this is the convenience of having  an operator whose matrix elements
are of natural magnitude, in this case of order unity.
\bibitem{footuno}
Although the numerical results obtained in the present analysis are similar to those
in the two-flavor case \cite{CCGL}, our conclusions about the relevance of the $O_3$
operator is different.
\end{thebibliography}
\end{document}